\documentclass[preprint,prb,onecolumn,amsmath,amssymb,superscriptaddress]{revtex4}
\usepackage[pdftex]{graphicx}
\usepackage{mathrsfs}
\usepackage[sort&compress]{natbib}

\newcommand{\be}{\begin{equation}}
\newcommand{\ee}{\end{equation}}
\newcommand{\bea}{\begin{eqnarray}}
\newcommand{\eea}{\end{eqnarray}}

\newcommand{\SI}{Supplementary Information}

\newcommand{\ZEM}{{E_Z/E_C}}

\newcommand{\nzexp}{Checkelsky2008,Checkelsky2009,Zhang2009b,Du2009,Bolotin2009,Zhang2010a}
\newcommand{\nzth}{Khveshchenko2001,Gorbar2002,Abanin2006,Alicea2006,Gusynin2006,Fertig2006,Fuchs2007,Herbut2007,Abanin2007a,Jung2009,Nomura2009,Shimshoni2009,DasSarma2009a,Hou2010,Kharitonov2011}
\newcommand{\allth}{Nomura2006, Yang2006,Alicea2007,Goerbig2006,Sheng2007, Lukyanchuk2008,Abanin2010a,\nzth}

\renewcommand{\phi}{\varphi}
\renewcommand{\epsilon}{\varepsilon}

\begin{document}
\title{Spin and valley quantum Hall ferromagnetism in graphene}
\author{A.~F.~Young$^\dag$}
  \affiliation{Department of Physics, Columbia University, New York, New York 10027, USA}
\author{C.~R.~Dean$^\dag$}
  \affiliation{Department of Electrical Engineering, Columbia University, New York, New York 10027, USA}
  \affiliation{Department of Mechanical Engineering, Columbia University, New York, New York 10027, USA}
\author{L. Wang}
  \affiliation{Department of Mechanical Engineering, Columbia University, New York, New York 10027, USA}
\author{H.~Ren}
  \affiliation{Department of Physics, Columbia University, New York, New York 10027, USA}
\author{P. Cadden-Zimansky}
  \affiliation{Department of Physics, Columbia University, New York, New York 10027, USA}
\author{K.~Watanabe}
\author{T.~Taniguchi}
  \affiliation{Advanced Materials Laboratory, National Institute for Materials Science, 1-1 Namiki, Tsukuba, 305-0044, Japan }
\author{J.~ Hone}
  \affiliation{Department of Mechanical Engineering, Columbia University, New York, New York 10027, USA}
\author{K.~L.~Shepard}
  \affiliation{Department of Electrical Engineering, Columbia University, New York, New York 10027, USA}
\author{P.~Kim}
  \affiliation{Department of Physics, Columbia University, New York, New York 10027, USA}
\date{\today}

\maketitle

\textbf{Electronic systems with multiple degenerate degrees of freedom can support a rich variety of broken symmetry states.  In a graphene Landau level (LL), strong Coulomb interactions and the fourfold spin/valley degeneracy lead to an approximate SU(4) isospin symmetry.  At partial filling, exchange interactions can break this symmetry, manifesting as additional Hall plateaus outside the normal integer sequence. Here we report the observation of a number of these quantum Hall isospin ferromagnetic (QHIFM) states, which we classify according to their real spin
structure using tilted field magnetotrasport. The large activation gaps confirm the Coulomb origin of all the broken symmetry states, but the order depends strongly on LL index. In the high energy LLs, the Zeeman effect is the dominant aligning field, leading to real spin ferromagnets hosting Skyrmionic excitations at half filling, whereas in the `relativistic' zero LL, lattice scale interactions drive the system to a density wave.}

\begin{figure*}[ht!]
	\begin{center}
	\includegraphics[width=\linewidth,clip]{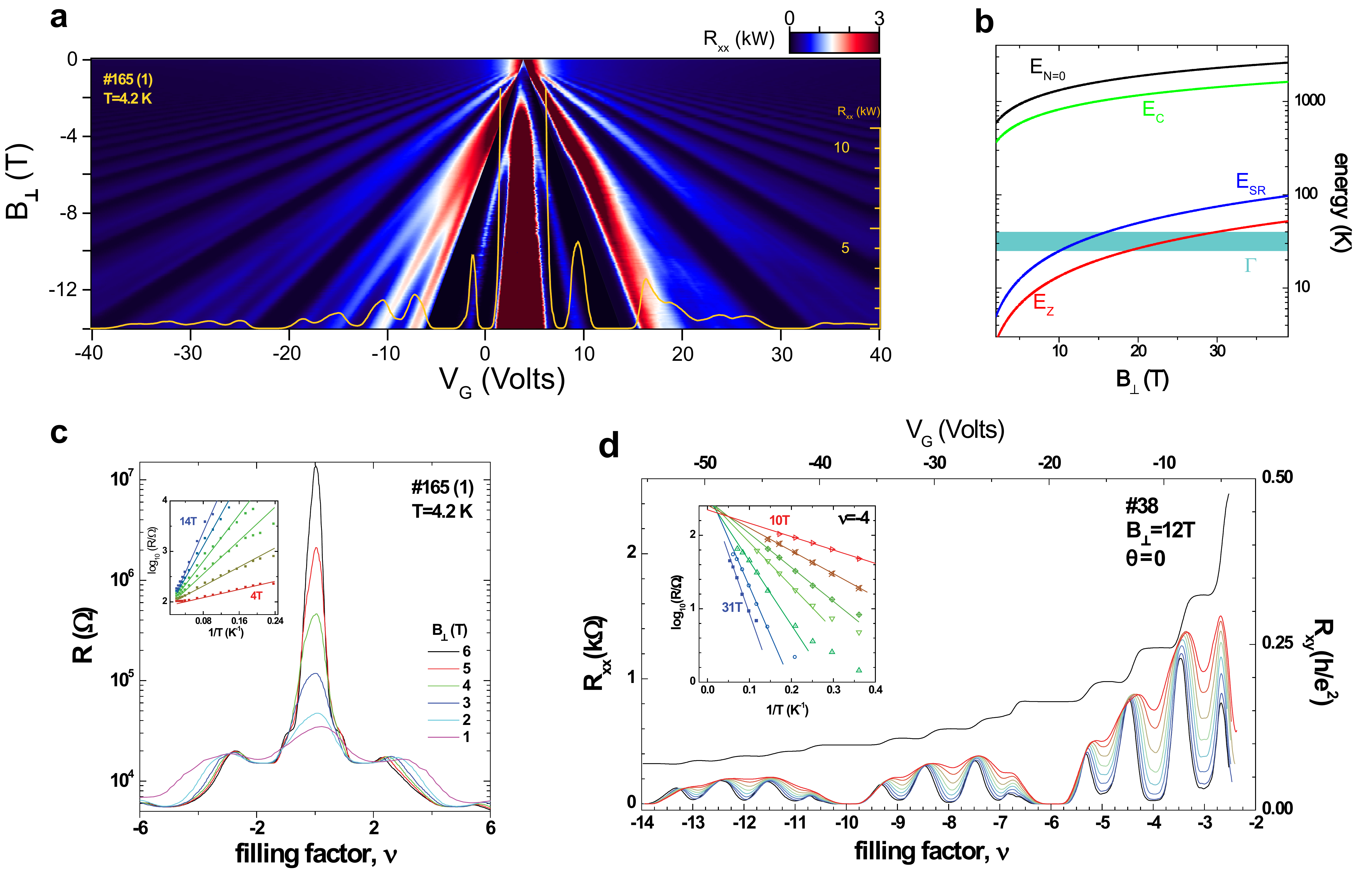}
		\caption{\textbf{All integer quantum Hall effect in graphene on h-BN.} \textbf{a} Landau fan from a monolayer graphene on h-BN device. Symmetry breaking of the LLs is visible from a few tesla, and at $B_\perp=14$T (superimposed) all integer filling factors feature minima in $R_{xx}$. The color scale for $-2<\nu<2$ has been expanded by a factor of 7.  \textbf{b} Energy scales.  The cyclotron gaps ($E_N\simeq\frac{\hbar v_F\sqrt{2}}{\ell_B}
$) and Coulomb energy ($E_C=\frac{e^2}{\varepsilon\ell_B}$) both characterize physics that does not distinguish between isospin flavors. The leading isospin anisotropies, which include the Zeeman effect ($E_Z=g\mu_B B_T$), lattice scale interactions ($E_{SR}\sim (a/\ell_B)E_C$)\cite{Alicea2006}, and disorder broadening($\Gamma$), are at least one order of magnitude smaller. Here, $v_F\simeq10^8\textrm{cm/s}$ is the Fermi velocity in graphene, $a=2.46 \AA$ is the lattice constant, $g_0=2$ is the bare gyromagnetic ratio, $\mu_B$ is the Bohr magneton, and  $\ell_B\sim$ 26 nm$/\sqrt{B_{\perp}\textrm{[Tesla]}}$ is the magnetic length.  The disorder energy scale is extracted from the magnetic field dependence of the Shubnikov-de Haas oscillations at low magnetic field in one representative device (see \SI).  \textbf{c} Development of the $\nu=0$ insulating state. Inset: temperature dependence showing the Arrhenius behavior of the insulating resistance. \textbf{d} Temperature dependence of the $R_{xx}$ minima in the symmetry broken IQHE regime. Inset: Arrhenius plots for $\nu=4$ as a function of magnetic field.}
		\label{fig1}
	\end{center}
\end{figure*}

The low energy effective theory of nearly neutral graphene describes two flavors of massless Dirac quasiparticles centered on the two inequivalent corners of the Brillouin zone, termed valleys. In a high magnetic field, the valley degeneracy combines with the physical electron spin to produce four component Landau levels, leading to the anomalous graphene quantum Hall sequence
\be \sigma_{xy}=\pm\frac{4e^2}{h}\left(N+\frac{1}{2}\right),\label{eq1}
\ee
where the LL index $N$ is a nonnegative integer and the additional factor of $1/2$ is related to the pseudospin winding number\cite{Geim2007}. The spin/valley degeneracy makes graphene a prime candidate for observing the rich physics of associated with multicomponent quantum Hall effects~\cite{Ando1982,EzawaBook,Shayegan2006}.  Graphene is exceptional as compared with its semiconductor counterparts due to a near-perfect energetic hierarchy (See Fig. \ref{fig1}b). The energy scales characterizing cyclotron motion ($E_N$) and long range interparticle Coulomb interactions ($E_C$)---both of which reflect physics that is independent of spin or valley flavor---dwarf explicit spin and valley symmetry breaking effects. The combined four flavor degeneracy can therefore be thought of as that of a single SU(4) isospin\cite{Goerbig2011a,Barlas2011}. As in other multicomponent quantum Hall systems, exchange interactions can drive the system through a ferromagnetic instability\cite{Nomura2006}, in which the order parameter corresponds to a finite polarization in a specific direction within the SU(4) isospin space. At integer fillings within a partially filled quartet LL, this order parameter is predicted to lead to a finite gap for charged excitations and a robust quantum Hall effect for integers outside the sequence described in Eq. (\ref{eq1}). The precise SU(4) polarization for given experimental conditions depends on the interplay between anisotropies arising from the Zeeman effect, lattice scale interactions, and disorder. All of these anisotropies are small and experimentally tunable, allowing for the possibility of a variety of distinct ground states across experimentally accessible ranges of filling factors, magnetic fields, and realizations of disorder.

Previous studies have indeed reported the observation of QHE at several integer filling factors outside the normal sequence\cite{Zhang2006,Jiang2007a,Checkelsky2008,Du2009,Dean2010,Dean2011}; however, the nature of the (presumably broken symmetry) states leading to these plateaus remains a matter of intense theoretical debate\cite{\allth}. In the N=0 LL, most experimental \cite{\nzexp} and theoretical \cite{\nzth} work has focused on the strongly insulating behavior observed at $\nu$=0, corresponding to half filling of the zero energy LL, which has no analog in conventional two dimensional electron systems. The insulating state has been described variously as a spin-polarized valley singlet, a valley polarized spin singlet, or a lattice scale spin density wave, but experimental resolution of this discrepancy has been hampered by the absence of any probe of the spin or valley order. Even less is known about the symmetry breaking at $\nu$=$\pm$1\cite{Zhang2006,Jiang2007a,Alicea2006,Alicea2007,Abanin2007b,Sheng2007,Lukyanchuk2008} or throughout the $N$ $\neq$ 0 LLs\cite{Zhang2006,Jiang2007a,Du2009,Dean2011}. Due to the anomalous structure of the $N=0$ LL, in which the valley quantum number corresponds to a real-space sublattice, the symmetry broken states for $N=0$ may not resemble those for $N\neq0$; however, limitations on sample quality and geometry in SiO$_2$-supported and suspended devices, respectively, have precluded a comparative study.

In this article, we address these issues by studying the thermal activation gaps, $^\nu\Delta $, associated with the broken symmetry IQHE states in graphene devices fabricated on hexagonal boron nitride (hBN) substrates\cite{Dean2010}. This gap is associated with energy cost of the lowest lying charged excitations of the ground state. Owing to the atomic-scale confinement of the electronic wavefunctions to the plane of the graphene, all orbital effects related to electronic interactions depend only on the out of plane component of magnetic field ($B_\perp$) while spins respond directly to the total magnetic field ($B_T$) independent of its direction. Tilted field measurements of the $^\nu\Delta$ thus allow us to extract quantitative information about the net spin of charged excitations in the broken symmetry states.

Fig. \ref{fig1}a shows the evolution of the quantum Hall effect with magnetic field in a representative device (\#165 thermal cycle 1). Symmetry breaking at $\nu=\pm1$ is visible at fields of $B_\perp\gtrsim 5~T$, followed by the higher LLs at $B_\perp\gtrsim7~T$; by 14~T (overlaid), $\rho_{xx}$ minima are visible at all integer filling within the experimental range. In addition, an insulating state develops at $\nu=0$ (Fig. \ref{fig1}c) starting from 2-3~T, consistent with previous work on clean, suspended graphene \cite{Du2009,Bolotin2009}. We find that all broken symmetry $\rho_{xx}$ minima, as well as the $\nu=0$ insulator, show simply activated temperature dependence over a wide range of fields (Fig. \ref{fig1}c and d, insets; note that \ref{fig1}d is data from a different device), allowing us to extract the energy gap $^\nu\Delta$ as a function of perpendicular and total field.  The exceptional quality of the devices studied here allows the observation of all integer filling broken symmetry states at magnetic fields of a few tesla, allowing the Zeeman energy to be tuned across a wide range in experimentally realizable magnetic fields. Using this technique, we explicitly demonstrate the dependence of the isospin ferromagnetic order, for fixed relative filling, on LL index $N$. For $N\neq0$ a dominant Zeeman anisotropy leads to spin polarized ground states at half filling and valley textured excitations at quarter filling, while in the $N=0$ LL, the situation is reversed: the $\nu=0$ insulator is shown to be unpolarized, and real spin textures form the most favorable charged excitations at from the fully polarized $\nu=1$.

\begin{figure*}[t]
	\begin{center}
	\includegraphics[width=\linewidth]{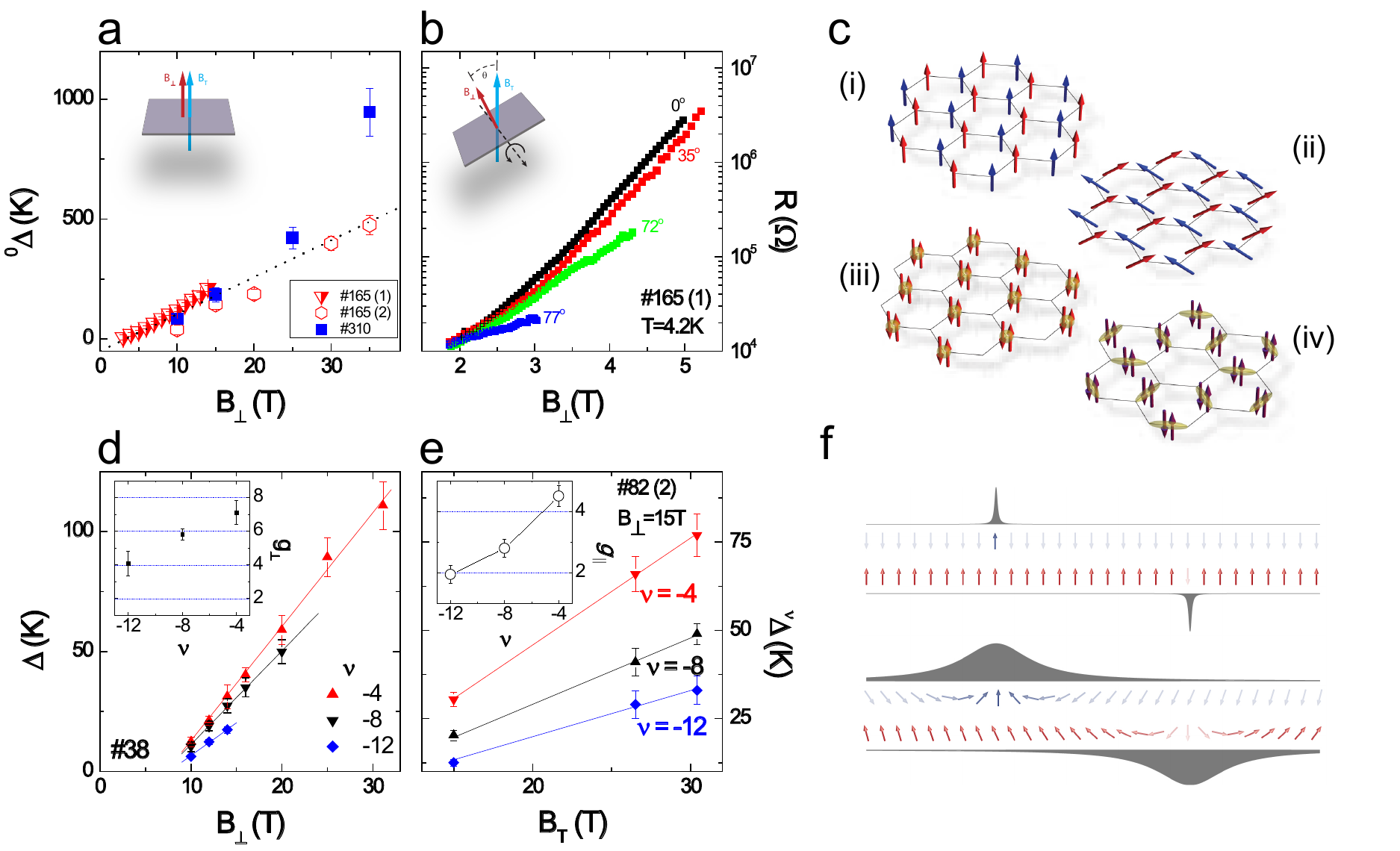}
		\caption{\textbf{a}
\textbf{Activation gaps of half filled quartet Landau levels.} \textbf{a} $B_\perp$ dependence of the $\nu=0$ gap, $^0\Delta$, for several devices. $^0\Delta$ increases approximately linearly with applied $B_\perp$, a feature not associated with any currently proposed theory for the $\nu=0$. The dashed line indicates $g_\perp=23$.  \textbf{b} Tilted field dependence of the resistance of the $\nu=0$ state. The resistance increases exponentially with field, consistent with a gapped state with $^0\Delta\propto B_\perp$. The resistance at fixed $B_\perp$ decreases for higher tilt angles, indicating a spin-unpolarized state. \textbf{c} Candidate QHIFM states for the $\nu=0$. Our experiment rules out the spin ferromagnet, (\textbf{i}); all other states are marked by lattice scale spin (for the canted antiferromagnet, (\textbf{ii})) or charge (for the charge density wave (\textbf{iii}) or Kekul\'{e} distortion (\textbf{iv})) order. \textbf{d} $B_\perp$ dependence of the half filled quartets for $N\neq0$, $\nu=-4,-8,-12$. Like the $\nu=0$, all gaps scale approximately linearly with $B_\perp$, with enhanced $g_\perp$ factors that decrease with increasing LL index. \textbf{e} Unlike the $\nu=0$ state, all activation gaps measured for half-filled LLs with $N\neq0$ increase with $B_T$, indicating spin polarized states. For $\nu=-4$ and $-8$, the enhancement of $g_\parallel$ indicates that charged excitations contain multiple flipped spins. \textbf{f} Schematic of charged excitations at half filling for $N\neq0$. Excitations into the spin-reversed conduction band can take the form of single reversed spin particle hole pairs or smoothly varying Skyrmion-antiSkyrmion (S-aS) spin textures, depending on the strength of exchange interactions relative to disorder and the Zeeman energy. At $B_\perp=15T$ in the samples studied in this work, the S-aS scenario prevails at $\nu=-4$ and $-8$, while charge at $\nu=-12$ is carried by single electron hole pairs.
}
		\label{fighalf}
	\end{center}
\end{figure*}

We start our discussion with the half filled LLs. Figs. \ref{fighalf}a and d show activation gaps for the half-filled LLs, $\nu$ = 0, -4, -8 and -12, as a function of $B_\perp$.  The perpendicular field dependence of $^\nu\Delta$  for all $N$ is qualitatively similar, following an approximately linear scaling with $B_\perp$.  We define the effective gyromagnetic ratio in \textit{perpendicular} field from the slope of these cures, $g_\perp\equiv \mu_B^{-1} \partial_{B_\perp} (^\nu\Delta)$.  We find that $g_\parallel$ is enhanced with respect to the bare value $g_0=2$.  The Coulomb energy is the only scale in the system compatible with the measured energy gaps, which are much larger than might be expected from known single-particle effects.  Moreover, $g_\perp$ decreases with increasing LL index $N$, consistent with exchange-driven quantum Hall ferromagnetism\cite{Nomura2006}.

Tilted field measurements reveal the uniqueness of the $N=0$ LL. For $N\neq0$, the half filled gaps, $^{-4}\Delta, ^{-8}\Delta,$ and $^{-12}\Delta$, increase with $B_T$ for fixed $B_\perp$ (see Fig. \ref{fighalf}d).  This is consistent with the existence of real spin polarized states, in which excitations involve quasiparticles containing a net spin reversal relative to the ground state (and applied magnetic field).  The activation gaps of such excitations consist of a direct Zeeman contribution from the reversal of spins against the external field, as well as an exchange contribution ($\Delta_X$) arising from the spin reversal relative to adjacent (polarized) spins,
\be \Delta=\Delta_X(B_\perp)+g_0\mu_B B_T-\Gamma.\ee
In contrast, the $\nu=0$ resistance decreases (Fig \ref{fighalf}\textbf{b} with increased $B_T$, an observation incompatible with the real spin polarized scenario for $\nu=0$\cite{Abanin2006,Fertig2006,Abanin2007a,Shimshoni2009}. Instead, the data suggest that a spin \textit{unpolarized} state, in which excitations contain a net spin aligned parallel (rather than antiparallel) to the applied field, underlies the insulating behavior at $\nu=0$.

Half filling of a fourfold degenerate graphene LL provides an ideal testing ground for the relative strength of the spin and valley anisotropies within the SU(4) isospin space. Because each cyclotron guiding center is doubly occupied, Pauli exclusion prevents the half-filled LL from fully polarizing in both spin and valley simultaneously.  As a result, spin- and valley- polarizing tendencies necessarily compete, and the resulting ground state reflects the result of that competition.  The fact that different order prevails at half filling for $N=0$ and $N\neq0$ even under identical experimental conditions ($B_\perp$ and $B_T$, which together fix the relative magnitude of the real spin anisotropy) suggests that the difference between LLs is intrinsic to graphene and originates in the valley sector.  A likely origin lies with the unique structure of the ZLL wavefunctions: whereas for the $N\neq0$ LLs wavefunctions in a single valley are spread equally over the two real space sublattices, for the ZLL electrons in a single valley are localized on a single sublattice \cite{Barlas2011}. Long range interactions do not distinguish between such lattice scale orbital structural difference, but \textit{short range} interactions do, potentially leading to different ground states in the $N=0$ and $N\neq0$ LLs\cite{Alicea2006}.  At $\nu=0$, the resulting interaction-induced valley anisotropies have been predicted to drive the system to one of a number of sublattice-ordered ground states ~\cite{Khveshchenko2001,Gorbar2002,Alicea2006,Gusynin2006,Herbut2007,Jung2009,Nomura2009,Hou2010,Kharitonov2011}, some of which are depicted in Fig. \ref{fighalf}c.  The experimental data presented here indicate that while the Zeeman effect wins the competition for the $N\neq0$ LLs, leading to spin polarized states at $\nu=-12,-8,$ and $-4$, the valley anisotropies dominate the zero LL, leading to the formation of one of the possible lattice scale density waves portrayed in Fig. \ref{fighalf}c (ii), (iii), and (iv).  The large size of the measure $^0\Delta$ gap, and its insensitivity to in-plane fields, suggest that the valley anisotropies may be $>10$ times stronger than their naive scale, $(1/\ell_B)\times E_C$, in line with renormalization group \cite{Aleiner2007,Kharitonov2011} and numerical calculations\cite{Jung2009}.

An additional notable feature of the experimental data is the linear scaling of $^0\Delta$ on $B_\perp$.  The absence of a spin-polarized ground state at $\nu=0$ precludes linear-in-$B_T$ Zeeman contributions to the excitation energy; meaning that the linear dependence must have an orbital origin.  In contrast to 2D electronic systems with parabolic dispersion, in monolayer graphene both $E_C$ and the cyclotron gaps scale as $\sqrt{B_\perp}$, a fact reflected in all available theories of the $\nu=0$ insulator based on the low-energy Dirac model. The observed linear scaling thus points to the relevance of physics beyond the low energy theory.  Motivating future work, quantitative data (See \SI) on the decrease of the gap in applied parallel field suggest that the spin ferromagnet, which is predicted \cite{Abanin2006,Fertig2006} to be an exact analog of the spin quantum Hall state\cite{Konig2007}, may be experimentally accessible in the best samples at high tilt angles in accessible magnetic fields ($B_T\lesssim$45~T).

\begin{figure}[t]
	\begin{center}
	\includegraphics[width=\linewidth,clip]{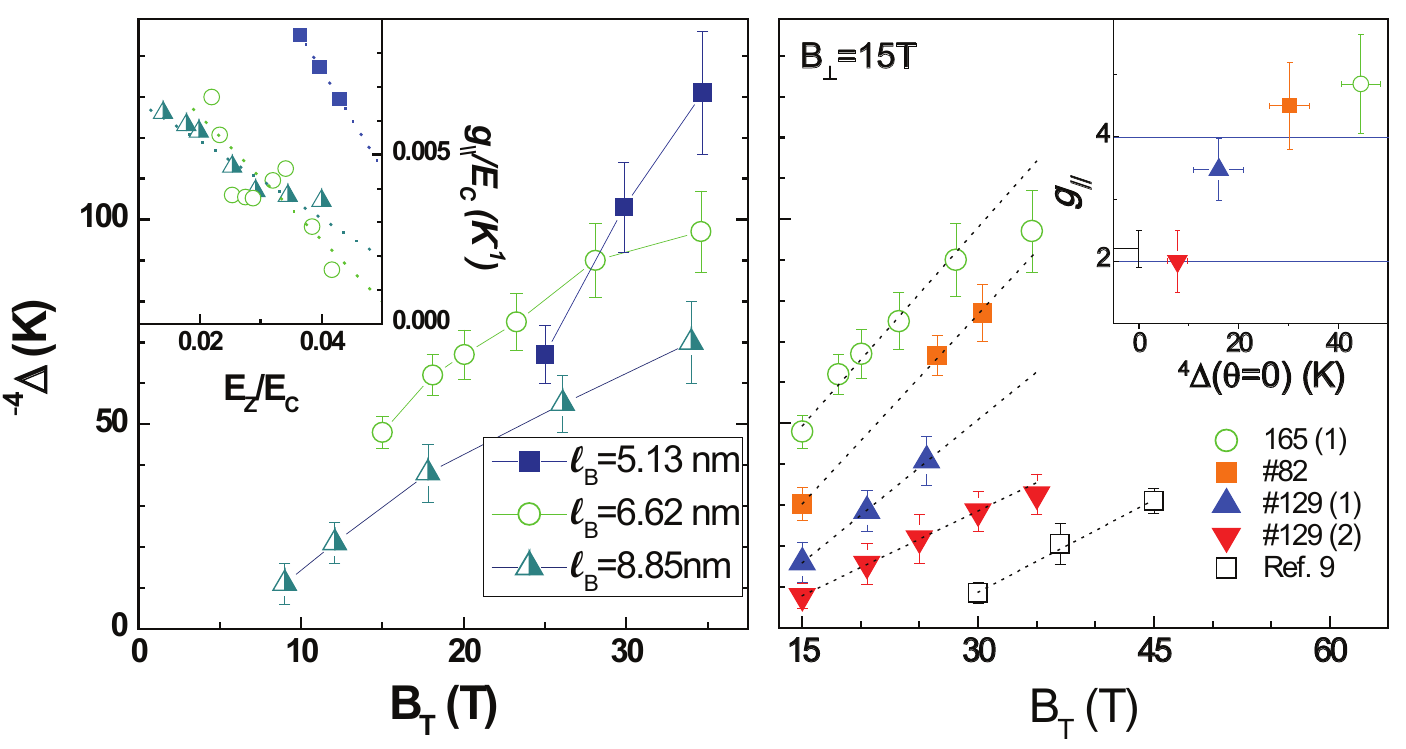}
		\caption{
\textbf{Skyrmion transport at $\nu= -4$.} \textbf{a} Perpendicular field dependence of $g_\parallel$ at $\nu=-4$. The three curves show tilted field dependence of $^{-4}\Delta$ for three different values of the $\ell_B$.  $g_{\parallel}$, calculated from nearest neighbor and next-to-nearest neighbor finite differences of the curves in the main panel, is negatively correlated with $\ZEM$ for fixed $\ell_B$.  For fixed $\ZEM$, $g_\parallel$ is negatively correlated with $\ell_B$, likely a combined effect of the increased exchange energy and decreased disorder parameter, $\Gamma/E_C$.
Lines are guides for the eye.
\textbf{b} Device dependence of $^{-4}\Delta$ for $B_\perp$=15T. Cleaner devices show both larger activation gaps (at all fillings) and larger $g_\parallel$ for fixed $\ell_B$.  This trend is observed both across devices and for the same device on different cooldowns, e.g. device \#129. The inset shows the device dependence of $g_{\parallel}$, which decreases with increasing disorder (for which $^{-4}\Delta$($\theta$=0)serves as a proxy variable).}
\label{figsky}
	\end{center}
\end{figure}

Despite the role of the single-particle Zeeman effect in setting the order in the higher LLs, tilted field activation gaps demonstrate that the symmetry breaking can be thought of as being essentially spontaneous, with Zeeman functioning as a small aligning field.  The gaps at half filling for $N\neq0$ increase with total magnetic field (Fig. \ref{fighalf}e) faster than might be expected for single spin flips, as reflected by the enhanced measured values of $g_\parallel\equiv \mu_B^{-1} \partial_{B_T}\Delta$ (Fig. \ref{fighalf}e, inset).  In a Zeeman-dominated spin polarized state, charge transport occurs through the thermal activation of spin-reversed particle-hole pairs (Eq. \ref{eq1}). While exchange contributions to the energy gap (the first term in Eq. \ref{eq1}) lead to $g_\perp>g_0$~\cite{Nicholas1988}, this enhancement does not carry through to $g_\parallel$: changing $B_T$ with $B_\perp$ fixed results in a measurement of the net spin of the excitation, and thus $g_\parallel=g_0$. In contrast, in exchange dominated spin polarized states it can be more energetically favorable to flip multiple spins smoothly in a Skyrmionic spin texture \cite{Sondhi1993,Fertig1994}, leading to a modified gap equation
\be \Delta=\Delta_X(B_\perp,K)+(2K+1)g_0\mu_B B_T-\Gamma\label{skygap}\ee
where $K\geq0$, the additional flipped spins per charged excitation, depends on the ratio $\ZEM$.  The observed $g_\parallel>2$ enhancements imply the presence of $K>0$ Skyrmionic charge carrying excitations at $\nu=-4$ and $\nu=-8$.

The widely tunable field effect natural to graphene facilitates the study of transport phenomena over a wide range of density in the same sample, allowing a detailed analysis of the dependence of $g_\parallel$ on LL index and $B_\perp$.  Our data is qualitatively in line with theoretical expectations for Skyrmion transport.  The decreasing trend in $g_\parallel$ with increasing LL index reflects the trend in exchange energy\cite{Yang2006}.  In contrast to two dimensional electronic systems with parabolic dispersion, where Skyrmions are expected only in the lowest energy LL\cite{Schmeller1995,Sondhi1993,Fertig1994}, Skyrmions are theoretically possible for all $|$N$|$$\leq$3 in monolayer graphene\cite{Yang2006} for vanishing E$_Z$.  We observe signatures of Skyrmions up to N=2, consistent with the presence of a finite Zeeman effect. Measurements of $g_\parallel$ at fixed filling for different values of $B_\perp$ in a single sample (see Fig\ref{figsky}a) show the expected decrease in $g_\parallel$---for fixed E$_C$---with increasing E$_Z$: the Zeeman coupling punishes large Skyrmions, driving down $K$ and, consequently, $g_\parallel$.  However, our data do not conform to the clean Skyrmion model, which predicts a universal scaling of the Skyrmion energy with the dimensionless parameter $E_Z/E_C$.  The inset of figure \ref{figsky}a shows the normalized in-plane dielectric constant, $g_\parallel/E_C$, for three different values of $B_\perp$ as a function of $E_Z/E_C$.  While the two lower $B_\perp$ curves collapse onto one another, in accordance with theoretical expectations for the clean limit, the $B_\perp$=25T data deviates strongly, implying the Skyrmions are larger than predicted\cite{Sondhi1993,Fertig1994,EzawaBook}.  The failure of a universal scaling with $E_Z/E_C$ suggests that a second dimensionless parameter, for example related to the disorder strength (e.g., $\Gamma/E_C$) may be relevant for Skyrmion transport\cite{Nederveen1999}.

We investigate the role of disorder in more detail by comparing several samples of varying quality.  Overall, the values of $g_\parallel$ observed in our graphene samples are lower than those measured---for similar values of $\ZEM$---in lower disorder GaAs quantum wells\cite{Schmeller1995}.  Moreover, we find that $g_\parallel$ is sample quality dependent.   Figure \ref{figsky}a shows tilted field dependence of $^{-4}\Delta$ at $B_\perp$=15T for several devices with different amounts of disorder.  While higher disorder is correlated with smaller LL gaps, consistent with an increased LL broadening $\Gamma$, disorder also leads to a smaller measured $g_\parallel$.  In the most disordered samples measured here, we find that $g_\parallel\approx2$, matching a previous measurement of SiO$_2$-supported graphene\cite{Zhang2006} and implying that in disordered samples, current is carried by single spin reversed particle-hole pairs.  We speculate that Skyrmion size may by principally limited not only by the Zeeman effect but by the disorder landscape, which may tend to favor smaller Skyrmion size\cite{Nederveen1999}.  The interplay between disorder and exchange may also contribute to the scaling of $g_\parallel$ with $B_\perp$ (Fig. \ref{figsky}a), which served to reduce the Skyrmion radius, $r_{sky}\propto (2K+1)\ell_B$\cite{EzawaBook}, in comparison to the length scale characterizing the disorder potential landscape.

\begin{figure}[t]
	\begin{center}
	\includegraphics[width=\linewidth]{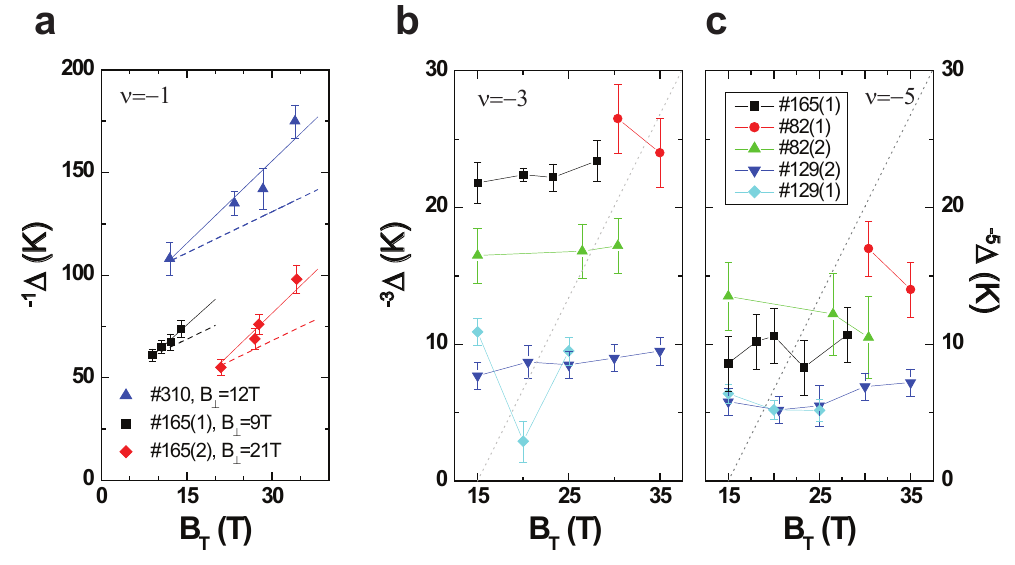}
		\caption{
\textbf{Transport phenomena at quarter filling.} \textbf{a} $\nu=-1$. Energy gaps increase with $B_T$, suggesting that excitations from the expected spin polarized, sublattice polarized state\cite{Alicea2006,Alicea2007,Sheng2007} involve real spin flips. Dependence on $B_T$ appears to support enhancement of $g_\parallel$ (solid lines are best linear fits to the data; dashed lines show $g_\parallel=2$ for reference). (b-c) Tilt dependence of $\nu=-3$ and $\nu=-5$. Most samples show minimal dependence on $B_T$, consistent with theoretical predictions of valley-textured excitations\cite{Yang2006}.  It is likely that in the absence of a single particle coupling field, these valley Skyrmions may be large \cite{Maude1996,Shkolnikov2005}.  Note reentrant behavior in sample \#129 (see Fig. \ref{figreen}). Dashed lines show $g_\parallel=2$ for comparison.}
		\label{figquart}
	\end{center}
\end{figure}

Like half filling, the phenomenology of a quarter filled graphene Landau level also shows markedly different behavior for $N=0$ and $N\neq0$. At quarter filling the naive ground state is a fully polarized state in which a single spin-valley flavor is occupied. While spin is always polarized in the direction of the field, valley anisotropies are thought to lead to Ising or x-y type valley polarizations for $N=0$ and $N\neq0$, respectively\cite{Alicea2007,Sheng2007}. Unlike at half-filling, spin and valley anisotropies do not compete with each other in the formation of the ground state at quarter filling: for a singly occupied cyclotron guiding center, there is no Pauli exclusion restriction on simultaneous spin- and valley polarization.  The interplay between spin- and valley- anisotropies does, however, contribute to the energetic of the excitation spectrum relevant for charge transport.  In the case of a dominant Zeeman effect, for example, the low lying charged excitations are thought to consist of valley flip textures~\cite{Yang2006} due to the high relative energetic cost of flipping real spins against the physical field. Most (although not all) activation measurements taken at quarter filling for $N\neq0$ (Figure \ref{figquart}b and c) are consistent with this scenario, with the gaps independent of $B_T$ for fixed $B_\perp$ to within experimental error.  In contrast, gaps at $\nu=-1$ increase with increasing $E_Z$ (Fig. \ref{figquart}a), suggesting again that the Zeeman effect is not the dominant anisotropy in the zero energy LL. Tilted field dependence of the gaps at $\nu=-1$ also show enhancement of $g_\parallel$ over the bare value, consistent with the large strength of exchange in the zero LL\cite{Yang2006}.

\begin{figure}[t]
	\includegraphics[width=\linewidth,clip]{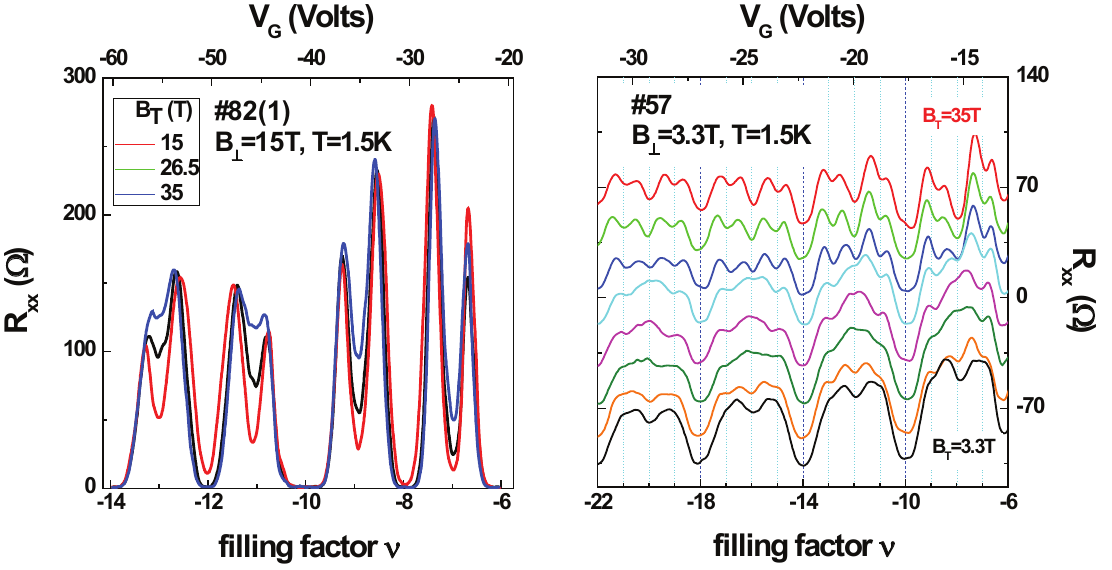}
		\caption{\textbf{Reentrant QHE in tilted field in the higher LLs.} (a) Gap collapse at odd filling. Gaps at quarter filling in the $N\neq0$ LLs collapse in applied in-plane field in selected samples. Gap collapse, or collapse and reemergence (cf. $^(-3)\Delta$ in device \#129 (1), Fig. \ref{figquart}) is device dependent. In the two devices in which it was observed, the behavior did not survive a subsequent cooldown. \textbf{b} At very low perpendicular magnetic fields, an additional minimum is visible only at half filling. With the addition of an in-plane field, this minimum disappears, before reemerging again as $B_T$ is increased further. Simultaneously, minima at quarter filling emerge with increasing $B_T$, indicating that they too can be spin active, increasing with $B_T$.}
		\label{figreen}
\end{figure}

While most data taken at half and quarter filling fit into the picture of QHIFM with LL index dependent anisotropy, we have observed several unexpected tilted field anomalies at both odd and even filling in the high LLs.  For example, among samples for which $^{-3}\Delta$ was measured as a function of $E_Z$, one (\#129 (1)) shows reentrant behavior, with the gap collapsing with increasing Zeeman and then growing again as $E_Z$ is further increased. Qualitatively similar behavior involving gap collapse in parallel field was observed in other samples throughout the quarter filled $N\neq0$ LLs, (see Fig \ref{figreen}b).  Reentrant  behavior was also observed at half filling in the higher LLs at low values of $B_\perp$ in one particularly high mobility sample (see Fig. \ref{figreen}b).  The origin of this unexpected behavior is not known, and the phenomenology is sample dependent: while the sample dependence makes the reentrant behavior unlikely to be an intrinsic property of clean graphene, it tends to appear in cleaner samples, and disappears upon sample contamination.  Among possible explanations are the unique susceptibility of the odd filling factors, to disorder\cite{Alicea2007,Abanin2007b}, which can stabilize spatially inhomogenous ground states, or an effect of the hBN dielectric, which may induce a superlattice-like potential and sample dependent spatially inhomogeneous valley anisotropies.

In summary, we have observed interaction driven spontaneously broken symmetry states at all integer filling factors in monolayer graphene.  The picture that emerges is one of exchange driven quantum Hall ferromagnetism within the combined spin-valley isopsin space\cite{Nomura2006}, where the leading anisotropies differ between the N=0 and N$\neq$0 LL.  Several questions remain, however: first, the precise nature of the $\nu=0$ state remains elusive, and the linear $B_\perp$ dependence is unaccounted for theoretically.  Moreover, the absence of spin polarization at $\nu=0$ and the presence of spin-reversed excitations at $\nu=1$, need to be reconciled with the prevailing theoretical models of the \textit{fractional} quantum Hall effect graphene $N=0$ LL~\cite{Papic2010,Toke2011}, all of which neglect the role of the valley anisotropies.   Finally, little is understood about the structure and excitation spectrum of the odd filling states, and in particular the anomalous tilted field dependence observed in certain samples.  We expect that the preliminary results presented here should motivate future work combining transport and surface science techniques, such as controlled absorption and scanned probe microscopy, to both elucidate the properties of these correlated states and, more generally, to use the graphene QHIFM as a model material platform for the systematic study of interacting systems in which all relevant experimental parameters, including disorder, can be tuned and probed \textit{in situ}.

\section{Methods}

We follow the fabrication method described in references \onlinecite{Dean2010,Dean2011} to produce multiterminal Hall bar and van der Pauw graphene devices on hexagonal-Boron Nitride (h-BN) substrates. Most devices presented were etched into the Hall bar geometry using a short exposure to $O_2$ plasma; all samples were annealed in a 5\% H$_2$/95\% Ar atmosphere to at 350$^{o}$ C for several hours to remove processing residues. Typical sample sized ranged from $.5\times.5$ to $5\times5$ $\mu$ m.

Devices were measured in a sample-in-$^4$He vapor variable temperature cryostat fitted with a mechanical sample rotation stage, mounted in the bore of a 35~T resistive magnet at the National High Magnetic Field Lab in Tallahassee, FL.  Electrical measurements for $\nu\neq0$ were performed in the four point geometry using a 10-100 nA current bias. For the $\nu=0$ measurements were taken in the two terminal geometry using a 200 $\mu V$ excitation voltage. The numerous features present in a gate voltage trace at intermediate magnetic fields ($B_\perp<25$~T) allowed precise angle calibration, with $B_\perp$ determined to better than .5\% accuracy. This was particularly important in the case of the $\nu=0$ state, where the dependence on $B_\perp$ is at least one order of magnitude stronger than that on $B_T$.

$R_{xx}$ minima were determined by sweeping the gate voltage at fixed temperature. All $\rho_{xx}$ minima, as well as the resistance maximum at $\nu=0$, obey an Arrhenius law, $R_{xx}\sim R_0\exp\left(\frac{-\Delta}{2T}\right)$. Gaps were determined by fitting to this formula over at least one decade of resistance when possible.  Error bars are dominated by ambiguity in picking out the appropriate `linear regime'. Plots for all gaps presented in the main text, including best fits, are available in the Supplementary information (SI)

Under normal conditions, graphene on h-BN samples are stable, and activation gaps can be measured repeatedly in the same sample over multiple cooldowns separated by weeks or months. However, rapidly warming the sample chamber causes outgassing from the cryostat walls, leading to the adsorption of debris on the graphene surface and higher disorder. This process, while difficult to control and not reliably reversible, allows us to exclude device-specific effects stemming, for example, from the interplay between the graphene electrons and the staggered lattice scale potential generated by the h-BN substrate.

\bibliography{references}
\bibliographystyle{naturemag1}

\noindent
$^{\dag}$These authors contributed equally to this work.
\bigskip

\noindent
Reprints and permission information is available online at http://npg.nature.com/reprintsandpermissions/. Correspondence and requests for materials should be addressed to PK.

\section{Acknowledgements}
We acknowledge discussions with I. Aleiner, A. Macdonald, Y. Barlas, R. Cote, W. Luo, and M. Kharitonov.

The measurements were performed at the National High Magnetic Field Laboratory, which is supported by National Science Foundation Cooperative Agreement No. DMR-0654118, the State of Florida and the US Department of Energy. We thank S. Hannahs, T. Murphy, and S. Maier for experimental assistance at NHMFL. This work is supported by DARPA CERA, AFOSR MURI, FCRP through C2S2 and FENA, NSEC (No. CHE-0117752) and NYSTAR. P.K. and A.F.Y. acknowledge support from DOE (DE-FG02-05ER46215).

\section{Contributions}
AFY and CRD conceived the experiment and analyzed the data. AFY, CRD, LW, and HR fabricated the samples. AFY, CRD, LW, HR, PC-Z performed the measurements. AFY, CRD, and PK wrote the paper.  JH, KS, and PK advised on experiments.

\end{document}